\documentclass[pdflatex,sn-nature]{sn-jnl}

\usepackage{graphicx}%
\usepackage{multirow}%
\usepackage{amsmath,amssymb,amsfonts}%
\usepackage{amsthm}%
\usepackage[title]{appendix}%
\usepackage{xcolor}%
\usepackage{textcomp}%
\usepackage{manyfoot}%
\usepackage{booktabs}%
\usepackage{algorithm}%
\usepackage{algorithmicx}%
\usepackage{algpseudocode}%
\usepackage{listings}%
\usepackage{natbib} %
\usepackage{booktabs} %
\usepackage{siunitx}  %
\usepackage{multirow} %
\usepackage{booktabs} %
\usepackage{siunitx}  %
\usepackage{rotating} %
\usepackage{xfrac}
\usepackage{indentfirst}

\usepackage{amsmath} %
\usepackage{cleveref} %
\usepackage{hyperref} %
\crefformat{equation}{Eq.~(#1)} %
\usepackage{lineno} %
\usepackage{cancel} %

\theoremstyle{thmstyleone}%
\theoremstyle{thmstyletwo}%

\theoremstyle{thmstylethree}%

\begin{document}

\title[Article Title]{White-box machine learning for uncovering physically interpretable dimensionless governing equations for granular materials}





\author[1]{\fnm{Xu} \sur{Han}}\email{hanxu8@connect.hku.hk}

\author[2]{\fnm{Lu} \sur{Jing}}\email{lujing@sz.tsinghua.edu.cn}

\author*[1]{\fnm{Chung-Yee} \sur{Kwok}}\email{fkwok8@hku.hk}


\author[3]{\fnm{Gengchao} \sur{Yang}}\email{yanggch8@mail.sysu.edu.cn}

\author[4]{\fnm{Yuri} \sur{Dumaresq Sobral}}\email{ydsobral@unb.br}

\affil*[1]{\orgdiv{Department of Civil Engineering}, \orgname{The University of Hong Kong}, \orgaddress{\city{Hong Kong}, \country{China}}}

\affil[2]{\orgdiv{Institute for Ocean Engineering}, \orgname{Shenzhen International Graduate School, Tsinghua University}, \orgaddress{\city{Shenzhen}, \postcode{518055}, \country{China}}}

\affil[3]{\orgdiv{School of Aeronautics and Astronautics}, \orgname{Sun Yat-sen University}, \orgaddress{\city{Guangzhou}, \postcode{510275}, \country{China}}}

\affil[4]{\orgdiv{Departamento de Matemática, Universidade de Brasília}, \orgname{Campus Universitário Darcy Ribeiro}, \orgaddress{\city{Brasília DF}, \postcode{70910-900}, \country{Brazil}}}


\abstract{Granular material has significant implications for industrial and geophysical processes. A long-lasting challenge, however, is seeking a unified rheology for its solid- and liquid-like behaviors under quasi-static, inertial, and even unsteady shear conditions. Here, we present a data-driven framework to discover the hidden governing equation of sheared granular materials. The framework, PINNSR-DA, addresses noisy discrete particle data via physics-informed neural networks with sparse regression (PINNSR) and ensures dimensional consistency via machine learning-based dimensional analysis (DA). Applying PINNSR-DA to our discrete element method simulations of oscillatory shear flow, a general differential equation is found to govern the effective friction across steady and transient states. The equation consists of three interpretable terms, accounting respectively for linear response, nonlinear response and energy dissipation of the granular system, and the coefficients depends primarily on a dimensionless relaxation time, which is shorter for stiffer particles and thicker flow layers. This work pioneers a pathway for discovering physically interpretable governing laws in granular systems and can be readily extended to more complex scenarios involving jamming, segregation, and fluid-particle interactions.}


\maketitle
\clearpage
\section{Introduction}\label{sec1}
Granular material, despite its enormous significance in the industrial and geophysical worlds, has a long-standing issue of missing complete governing equations. The missing piece is a unified rheology that captures its solid- and liquid-like behaviors when subjected to quasi-static, inertial, and even unsteady shear conditions~\cite{vo2020additive,mandal2020insights,guillard2016scaling,wu2025unified,aguilar2016robophysical,berzi2024granular}. The main challenge stems from the significant influence of multiple interacting parameters on granular rheology, including loading stress, solid fraction, flow geometry, boundary roughness, particle shapes, particle size distributions and interactions with interstitial fluids \cite{kamrin2024advances,kim2020power,zhao2023role,yao2021competing}. The complexity of transient flows is further intensified due to their inherent instabilities. Traditionally, researchers have individually analyzed these parameters through numerical simulations \cite{dartevelle2004numerical} and laboratory tests \cite{gdr2004dense}, a process that relies heavily on human expertise and is highly time-consuming. During the past few decades, the most successful constitutive model for flowing granular materials has been the $\mu(I)$ rheology \cite{gdr2004dense,jop2006constitutive,forterre2008flows}, in which the bulk friction coefficient $\mu=\tau/P$ (the ratio of shear stress $\tau$ to confining normal pressure $P$) is governed by the dimensionless inertial number $I$. 
 This inertial number is formally expressed as $I=\dot{\gamma}d/\sqrt{P/\rho}$, where $\dot{\gamma}$ denotes the shear rate, and $d$ and $\rho$ represent the diameter and density of the grains, respectively. Despite recent advances that have broadened the applicability of $\mu(I)$ rheology to real-world scenarios \cite{blatny2024critical,fei2020particle,bouchut2021dilatancy,caucao2025nonlinear,yangEfficientLatticeBoltzmann2023,yangFrictionalBoundaryCondition2023} and facilitated its integration into large-scale continuum simulations \cite{liu2022including,fei2020simulation,gray2014depth}, this model does not adequately represent inhomogeneous fields at low shear rate ($I<10^{-3}$) where non-local effects are not negligible \cite{cheng2021unified}. In addition, transient processes for granular flows are beyond the scope of these steady-state rheological models. Thus, conventional research protocols struggle to discern a unified and non-stationary rheological model for granular assemblies under time-dependent loading conditions, necessitating the pursuit of a more efficient avenue for advancing the research.

Artificial intelligence (AI) has experienced exponential progress over the past decade, propelled by advances in computational power and the increasing availability of large datasets \cite{leng2023fifth,krizhevsky2012imagenet,lecun2015deep,jordan2015machine,wang2023scientific,jumper2021highly,abramson2024accurate,zhao2024artificial}. Numerous studies \cite{mahmoudabadbozchelou2022digital,bahiuddin2024review,dahl2025predicting,davoodi2023hybridized,chen2021recurrent} have shown that machine learning (ML) models, typified by various neural networks (NNs), excel at mapping complex rheological relationships for fluids without relying on expert knowledge. Nevertheless, these traditional ML models are purely data-driven and operate as black boxes, hence hindering our understanding of the underlying physics. In contrast to the black-box nature of traditional ML methods, revolutionary algorithms have emerged to inversely extract underlying physical laws and governing equations from empirical data \cite{brunton2016discovering,brunton2024promising,wang2025deep,chen2021physics}. These approaches are more readily applicable to interdisciplinary tasks and have become a leading trend in the so-called AI4Science field. Pioneering work by Lipson and co-workers \cite{bongard2007automated,schmidt2009distilling} introduced stratified symbolic regression and genetic programming to extract the underlying differential equations that govern nonlinear system dynamics from datasets. Subsequently, these genetic algorithms have gained significant developments, with notable advancements including AI Feynman \cite{udrescu2020ai}, AI-Descartes \cite{cornelio2023combining}, and the open-source library PySR \cite{cranmer2023interpretable}. In parallel, another mainstream direction is the sparsity-promoting approach. The seminal breakthrough is known as the sparse identification of nonlinear dynamics (SINDy) for ordinary differential equations (ODEs)~\cite{brunton2016discovering}, which has subsequently evolved into the PDE-Find method to identify partial differential equations (PDEs) \cite{rudy2017data}. SINDy and PDE-Find utilize sparse regression to select dominant candidate functions from a high-dimensional nonlinear function space, thereby revealing parsimonious governing equations. The sequential threshold ridge regression (STRidge) technique achieves sparsity by iteratively computing sparse solutions with hard thresholds, thus striking a balance between the accuracy and complexity of the identified equations. Compared to symbolic regression, SINDy is more efficient and robust in high-dimensional spaces despite requiring some prior knowledge to design the candidate library~\cite{brunton2016sparse,boninsegna2018sparse,kaheman2020sindy,shea2021sindy,eganAutomatically2024}.

Recently, SINDy and its variants have been successfully applied in the rheological community. For example, an unbiased rheological relation for soft polymer microgel materials has been successfully identified from experimental data, taking the form of a differential equation that governs the time derivative of stress~\cite{mahmoudabadbozchelou2024unbiased}. Another notable development is the Rheo-SINDy framework, proposed to derive a constitutive model for viscoelastic fluids directly from rheological data~\cite{sato2025rheo}. In particular, this framework highlights that oscillatory shear tests are more suitable than constant shear tests for generating training data in equation discovery. These works provide valuable inspiration for our data-driven investigation into granular rheology. Nevertheless, applying current data-driven methods in the SINDy paradigm to granular systems faces critical challenges: (i) the high-noise granular data, arising from the discrete nature of particle interactions, render precise numerical differentiation difficult; (ii) the derived governing equations are case-specific with non-meaningful coefficients, hindering generalization and interpretability of granular micro-mechanisms. 

This study introduces an advanced white-box framework called PINNSR-DA that seamlessly synergizes the robust representation learning capabilities of deep neural network (DNN), the precision of automatic differentiation for derivative calculations, and dimensional analysis for deciphering equation coefficients. Leveraging this framework, we successfully identify a consistent dimensionally homogeneous differential equation for both steady-state and transient granular flows in quasi-static regime, demonstrating remarkable generalization ability across various loading scenarios and configurations. Notably, we identify a dominant dimensionless number that governs the equation coefficients and relates to the relaxation time of granular system. This enhances the generalizability of the discovered equations, eases their application to larger scales, and yields novel insights into the underlying physics. Furthermore, we provide microscopic interpretations of each term in the discovered equation in terms of fabric, underscoring the reliability and accuracy of our model. The contributions of this paper are twofold: (i) it presents methodological innovation (PINNSR-DA) for deriving dimensionally homogeneous governing equations tailored for granular flows, and (ii) to the best of our knowledge, it marks the first successful application of data-driven discovery approaches to the field of unknown governing equations for granular flows, which extends beyond the scope of steady-state rheological model (like the $\mu(I)$ rheology) to capture the evolution of stress under time-varying loading conditions, thus expanding our understanding of granular physics and bridging a significant research gap in AI4Science for granular rheology.

\section{Results}\label{sec2}
\subsection{Data Collection}\label{sec:data_collection}
\begin{figure}[htbp]
    \centering
    \includegraphics[width=0.99\linewidth]{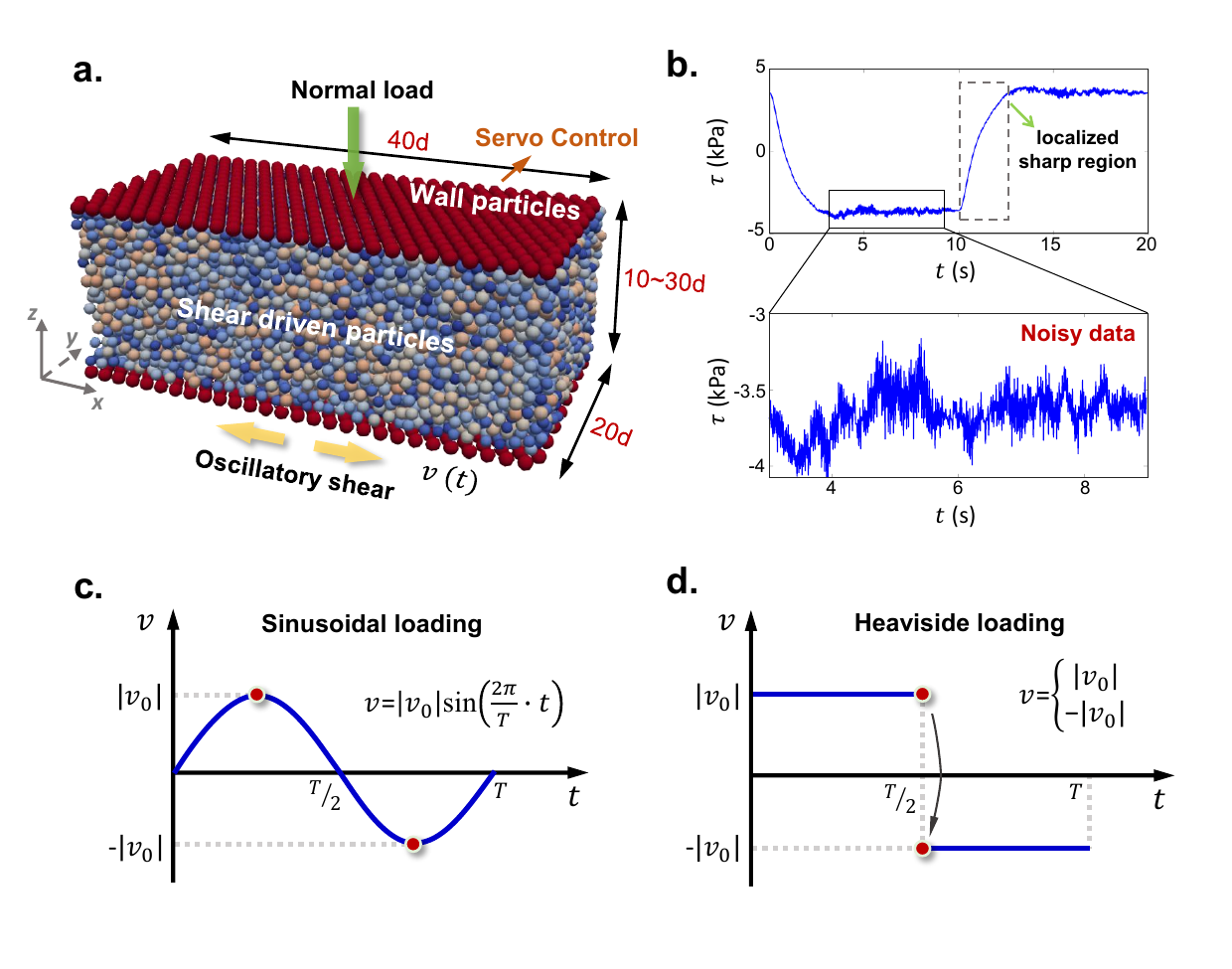}
    \caption{\textbf{Data collection by DEM simulations.} \textbf{a} Two-plate DEM simulation setup for transient granular flows. \textbf{b} Representative shear stress ($\tau$) measurement on the bottom layer under reversal loading, which exhibits significant noise due to the discrete nature of granular interactions. \textbf{c} Sinusoidal loading. \textbf{d} Heaviside loading.}
    \label{Fig_Data_collection}
\end{figure}

We perform discrete element method (DEM) simulations of wall-confined oscillatory shear tests on dry granular medium using LIGGGHTS \cite{kloss2012models}. This setup represents the simplest configuration that allows the analysis of both steady-state and transient responses for granular flows. As depicted in Fig.~\ref{Fig_Data_collection}a, granular assemblies with a mean diameter $d_p$ and $\pm10\%$ uniform polydispersity (to prevent crystallization) are sheared in a three-dimensional (3D) periodic domain. The domain dimensions are set to $L=40d_p$ in the stream-wise ($x$) direction, $W=20d_p$ in the span-wise ($y$) direction and variable height $H=10\sim30d_p$ along the $z$-axis. The upper and lower plates comprise similar particles that constrain relative motion, thus acting as rigid planes with adequate roughness, and gravity is nullified to prevent strain localization at the boundaries. Particle interactions follow the Hertz contact model, with Young’s modulus varying between $5\times10^7$ and $1\times10^9 $Pa, Poisson’s ratio of 0.4, particle-particle friction coefficient of 0.5, and restitution coefficient of 0.8. In addition, servo control is utilized to maintain a constant overburden pressure on the top wall while dynamically shearing is applied at the bottom wall by adjusting its velocity $v$ in the $x$ direction. 

We regulate $v(t)$ of the bottom wall to match two predefined functions to achieve transient reversal conditions: (i) sinusoidal loading and (ii) Heaviside loading, as illustrated in Figs.~\ref{Fig_Data_collection}c and~\ref{Fig_Data_collection}d, respectively. This approach imposes a time-dependent shear rate $\dot{\gamma}(t)$. Taking the sinusoidal case as an example, $\dot{\gamma}(t)$ adheres to the function described in Eq. (\ref{eq1_shear_rate}):

\begin{equation}
\dot{\gamma}=\frac{v_0}{H}\sin\left(\frac{2\pi}{T}t\right),\label{eq1_shear_rate}
\end{equation}
where $T$ and $v_0$ are the period and amplitude of sinusoidal velocity, respectively. By varying $v_0$, we can subject the granular materials to different shear conditions. For simplicity, this work focuses on rate-independent rheology, in which the amplitude of the shear rate ${\dot{\gamma}_0}= v_0/H $ is controlled to maintain the inertial number $I \leq 10^{-3}$.

This simulation facilitates the capture of the temporal evolution of stresses $\sigma_{ij}$ (where $i, j=x, y, z$) during the transient process. Herein, we primarily focus on normal stress ($\sigma_{zz}$) and shear stress ($\sigma_{xy}$), which will hereafter be denoted as $P$ and $\tau$, respectively. The DEM data are then used to train the data-driven framework, aiming to establish governing differential equations.  However, as illustrated by the representative curve in Fig.~\ref{Fig_Data_collection}b, the recorded signals exhibit substantial fluctuations due to random collisions and the discrete nature of granular interactions. Furthermore, there are localized sharp regions caused by shear reversal. Such noisy data with irregularities pose considerable challenges for numerical differentiation, a longstanding challenge for existing equation discovery approaches (e.g., SINDy).

\subsection{PINNSR-DA framework}\label{subsec2-2}

\begin{figure}[htbp]
    \centering
    \includegraphics[width=0.99\linewidth]{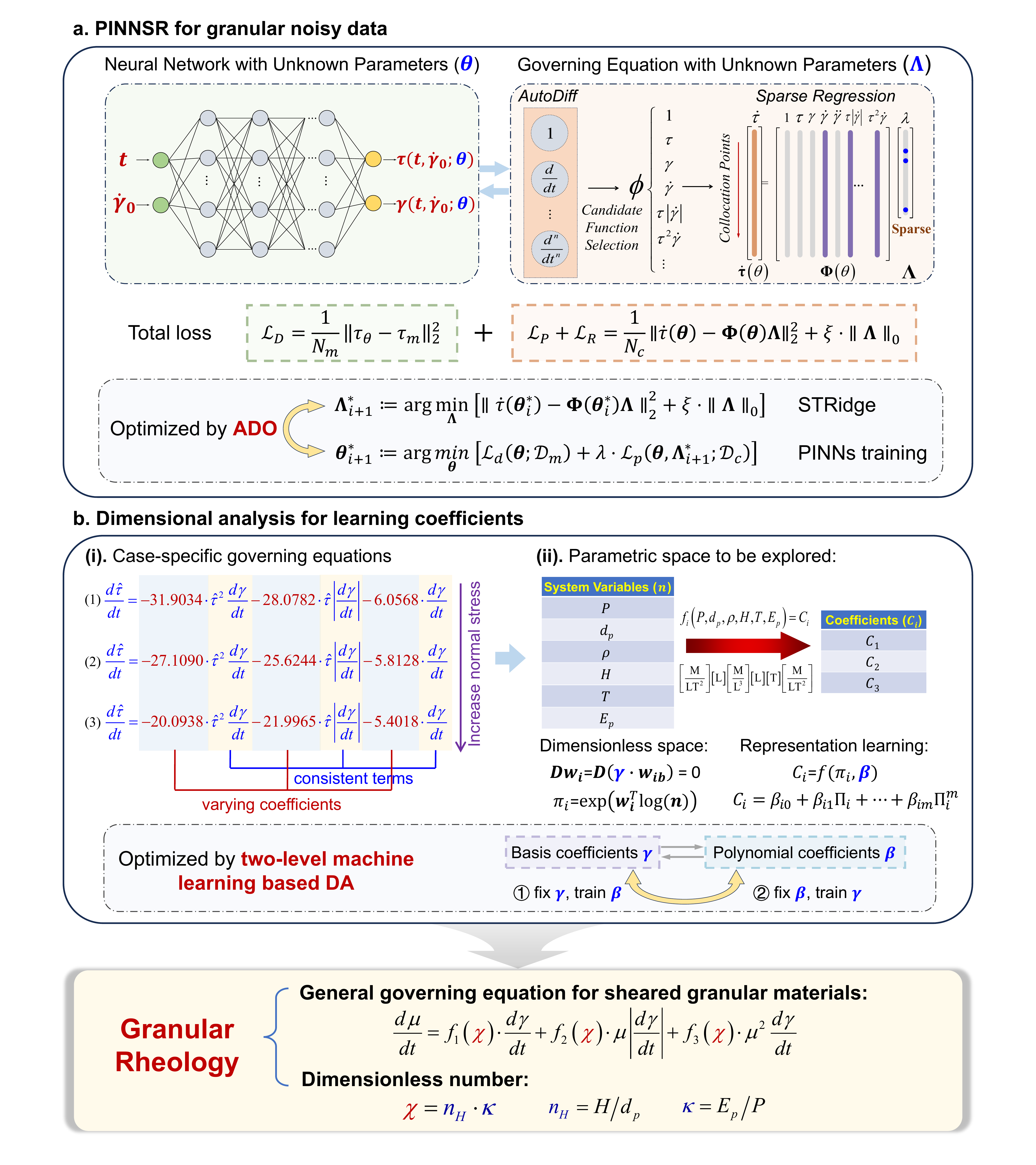}
    \caption{\textbf{PINNSR-DA: physics-informed neural networks with sparse regression and dimensional analysis.} \textbf{a} Physics-informed neural networks with sparse regression (PINNSR), optimized by alternating direction optimization (ADO) algorithm. \textbf{b} Dimensional analysis for learning coefficients: \textbf{(i).} case-specific governing equations identified by PINNSR for each configuration, exemplified by varying normal stress: (1) $P=1\times10^4$  Pa, (2) $P=1.5\times10^4$ Pa, and (3) $P=2.5\times10^4$ Pa; \textbf{(ii).} Parametric space to be explored via dimensional analysis for deriving explicit expressions of equation coefficients. The anticipated outcome of PINNSR-DA is a general governing equation for sheared granular materials, with coefficients dictated by physically meaningful dimensionless numbers.}
    \label{Fig_framework}
\end{figure}
We present a novel framework called PINNSR-DA to unveil a dimensionally homogeneous governing equation in granular systems (see Fig.~\ref{Fig_framework}). This framework
streamlines the identification of consistent parameterized governing equations into two steps: (i) the first step is to identify a case-specific equation by physics-informed neural networks with sparse regression (PINNSR), in which the regression coefficients vary across cases due to changes in system variables (e.g., granular properties, loading conditions, and flow geometry); (ii) the second step is to recover the expressions of the varying coefficients by machine learning-based dimensional analysis (DA).

We consider a general form of the rheological governing equation implicit in the noisy granular data, as given by Eq. (\ref{eq2_general_form}), where $\tau$ denotes the latent solution of the transient stress state under the applied strain $\gamma$:
\begin{equation}
\dot{\tau}=\mathcal{F}\left[\tau,\ \gamma,\dot{\gamma},\ddot{\gamma},\ldots,\tau\gamma,\tau\dot{\gamma},\tau^2\dot{\gamma},\ldots;\mathbf{\Lambda}\right].\label{eq2_general_form}
\end{equation}
Here, $\mathcal{F}(\cdot)$ represents a general nonlinear functional that encompasses $\tau$, $\gamma$, their temporal derivatives, and various combinations thereof, parameterized by $\mathbf{\Lambda}$. We hypothesize that $\mathcal{F}(\cdot)$ can be approximated using sparse regression, drawing on a few key terms selected from an extensive library of candidate functions. Subsequently, Eq.~(\ref{eq2_general_form}) can be reformulated as:
\begin{equation}
\dot{\tau}=\mathbf{\Phi}\mathbf{\Lambda},\label{eq3_sparse_regression}
\end{equation}
where $\mathbf{\Phi}\in\mathbb{R}^{l\times n}$ is a vast library of notational functions encompassing numerous candidate terms. Here, $l$ denotes the length of the candidate vector, aligned with the time derivative $\dot{\tau}$ on the left-hand side, and $n$ represents the total count of candidate terms within the library. To ensure that the discovered equations remain impartial to our prior knowledge regarding the system, we include as many diverse variants as possible, formed from shear stress, shear strain, and their time derivatives, into the candidate library ($\mathbf{\Phi}$). Meanwhile, $\mathbf{\Lambda}\in\mathbb{R}^{n\times1}$ is the sparse vector of coefficients where only the active candidate terms have nonzero values. The objective now reduces to the identification of a sparse $\mathbf{\Lambda}$ through sparsity-promoting methods, thereby establishing the formulation of $\dot{\tau}$.

In the PINNSR step (Fig.~\ref{Fig_framework}a), we draw inspiration from the seminal work~\cite{chen2021physics} and develop a PINNSR method for granular rheology that simultaneously predicts the system responses by deep neural network (DNN) and identifies ODEs in parsimonious closed forms based on a specific case. This method capitalizes on the automatic differentiation of DNN to accurately compute derivatives, providing precise $\mathbf{\Phi}$ for sparse regression (SR) from noisy granular data. Subsequently, it incorporates the SR-identified rheological equation as a physical constraint in the DNN loss function to curb overfitting. More specifically, the DNN utilizes dual inputs: temporal coordinate $t$ and shear rate amplitude ${\dot{\gamma}}_0$, with outputs of shear stress $\tau$ and strain $\gamma$. Here, ${\dot{\gamma}}_0$ is flattened into a vector of the same length as $t$, acting as an identifier for each loading case, which enables concurrent training across multiple loading conditions. This structure learns the latent solution by minimizing the loss function $\mathcal{L}_D\left(\boldsymbol{\theta};\mathcal{D}_m\right)$, where $\boldsymbol{\theta}$ denotes the trainable parameters (including weights and biases) and $\mathcal{D}_m$ represents the measurement data employed in training. In addition, the sparse representation of the reconstructed ODE can be cast in a residual form as $\mathcal{R}=\dot{\tau}-\mathbf{\Phi\Lambda}$. This residual is evaluated in the collocation set ($\mathcal{D}_c$) distributed over the temporal domain and is subsequently used to compute the physics residual loss function $\mathcal{L}_P(\boldsymbol{\theta},\boldsymbol{\Lambda};\mathcal{D}_c)$ by the mean squared error (MSE) metric. A residual-based adaptive sampling strategy is introduced that dynamically updates the distribution of $\mathcal{D}_c$ at each training iteration according to the residual $\mathcal{R}$ from the previous step. This strategy facilitates sampling more points in a sharp localized region with steep gradients, significantly improving the performance of PINNs. Afterwards, we formulate the total loss function as follows:
\begin{equation}
\mathcal{L}(\boldsymbol{\theta},\boldsymbol{\Lambda};\mathcal{D}_m,\mathcal{D}_c)=\mathcal{L}_D(\boldsymbol{\theta};\mathcal{D}_m)+\lambda\cdot\mathcal{L}_P(\boldsymbol{\theta},\boldsymbol{\Lambda};\mathcal{D}_c)+\xi\cdot\|\boldsymbol{\Lambda}\|_0,\label{eq4_PINNSR_loss}
\end{equation}
where $\boldsymbol{\theta}$ and $\boldsymbol{\Lambda}$ are trainable parameters for DNN and SINDy, respectively; $\lambda$ and $\xi$ are hyperparameters that regulate the weight of the residual physics loss and the penalty for equation complexity; and $\|{\cdot}\|_0$ denotes the $l_0$ norm. Upon minimizing the total loss function as defined in Eq. (\ref{eq4_PINNSR_loss}), the neural network is capable of fitting the measurement data while adhering to the constraints imposed by the underlying rheological ODE. To efficiently address the optimization of such high-dimensional problems, an alternating direction optimization (ADO) algorithm is utilized. This approach decomposes the overall optimization process into a series of manageable subproblems, as illustrated in Eqs.~(\ref{eq5_argmin1}) and (\ref{eq6_argmin2}):

\begin{equation}
\boldsymbol{\Lambda}_{i+1}^*:=\arg\min_{\boldsymbol{\Lambda}}\left[\left\|\dot{\tau}(\boldsymbol{\theta}_i^*)-\boldsymbol{\Phi}(\boldsymbol{\theta}_i^*)\boldsymbol{\Lambda}\right\|_2^2+\xi\cdot\left\|\boldsymbol{\Lambda}\right\|_0\right],\label{eq5_argmin1}
\end{equation}

\begin{equation}
\boldsymbol{\theta}_{i+1}^*:=\arg\min_{\boldsymbol{\theta}}\left[\mathcal{L}_d(\boldsymbol{\theta};\mathcal{D}_m)+\lambda\cdot\mathcal{L}_p(\boldsymbol{\theta},\boldsymbol{\Lambda}_{i+1}^*;\mathcal{D}_c)\right],\label{eq6_argmin2}
\end{equation}
thereby facilitating the learning of the trainable parameters in a structured and systematic manner. More details can be found in \emph{Methods} and \textcolor{red}{Section 1.1 of the SI}. Leveraging PINN-SR, we can derive precise governing equations for specific configurations.

In the DA step (Fig.~\ref{Fig_framework}b), we observe that changes in system variables lead to equations with consistent terms but varying coefficients (see Fig.~\ref{Fig_framework}b(i) for conceptual illustration). Therefore, an additional step is required to determine the form of these coefficients. We apply the dimensionless learning method cited in the notable work \cite{xie2022data} to discover the dominant dimensionless number and scaling laws of each coefficient. As depicted in Fig.~\ref{Fig_framework}b(ii), we utilize a two-level machine learning (ML) algorithm to estimate the parameters $\boldsymbol{\gamma}$ and $\boldsymbol{\beta}$, which determine the dimensionless number and the representation learning model, respectively. This optimization scheme alternates between updating $\boldsymbol{\beta}$ with $\boldsymbol{\gamma}$ fixed and vice versa, iterating until both sets of parameters converge. In this ML-based manner, we expedite the identification of the dominant dimensionless numbers with interpretability and extend the sparse $\boldsymbol{\Lambda}$ from constant values to expressions of system variables. Further details are provided in \emph{Methods} and \textcolor{red}{Section 1.2 of the SI}.

\subsection{Learning case-specific governing equation}\label{subsec2-3}
\begin{figure}[htbp]
    \centering
    \includegraphics[width=0.99\linewidth]{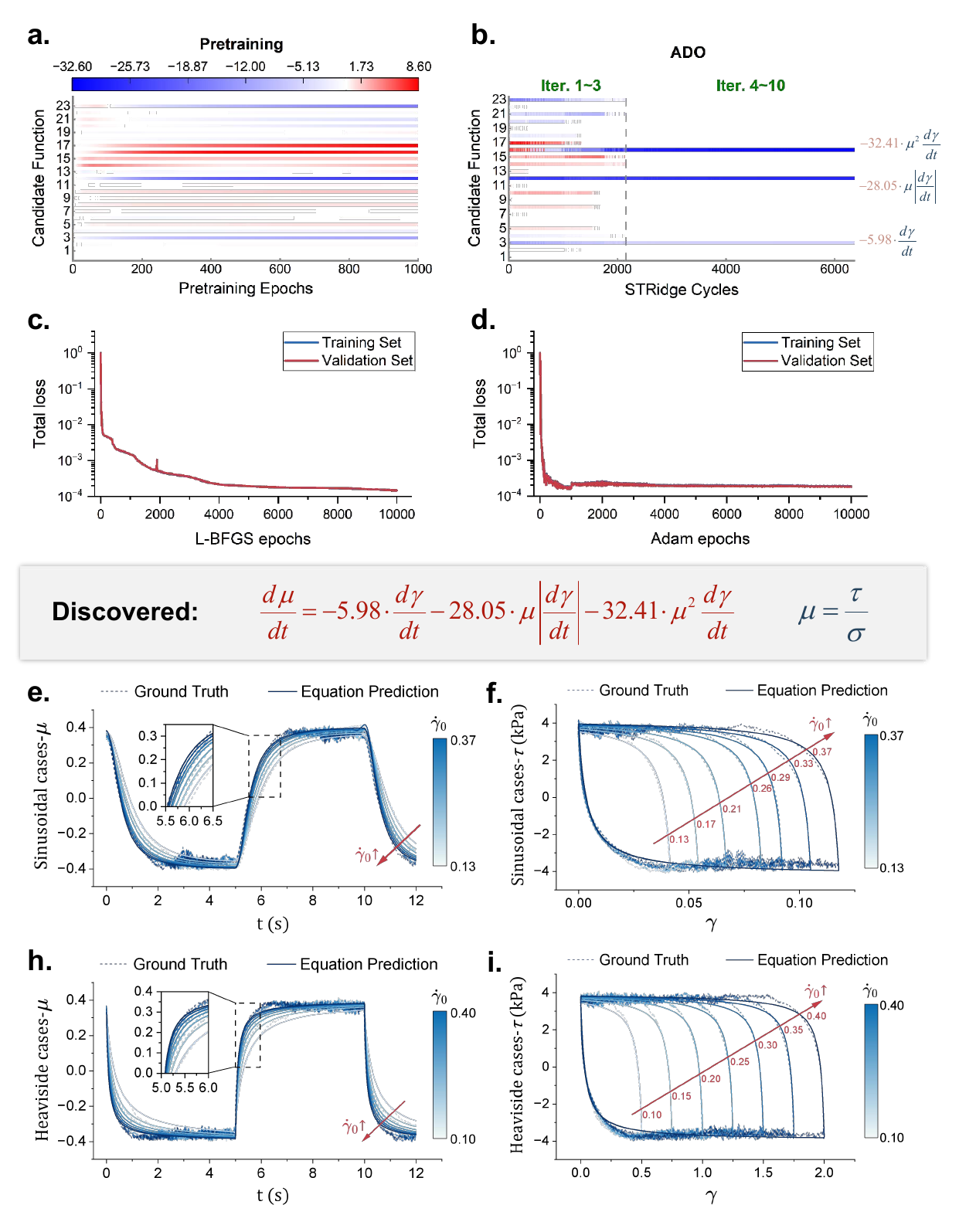}
    \caption{\textbf{Identification of governing equation: Training process, Equation form, and Validations.} \textbf{a-b} Evolution of the sparse coefficients for 23 candidate functions used to construct the governing equation, where the color represents the coefficient value. \textbf{c-d} Total loss for both the training and testing sets decreases progressively, ultimately reaching a stable and acceptable range. \textbf{e-f} Validations in prediction sets of Sinusoidal loading cases (dashed lines represent DEM results, while the solid lines correspond to the outcomes from the Runge-Kutta integration of the identified equation). \textbf{h-i} Validations of Heaviside loading cases.}
    \label{Fig_PINN_SR_results}
\end{figure}

This section presents the findings identified by PINNSR using an example of a case-specific governing equation. We examine a representative system characterized by parameters $P=10~\text{kPa}$, $H=15~d_p$, $T=10~\text{s}$, with particle properties $\rho=2500~\text{kg}/\text{m}^{3}$, $d_p=0.005~\text{m}$, and $E_p=5\times10^7~\text{Pa}$. Based on these DEM settings, we vary the amplitude of the shear rate ${\dot{\gamma}}_0=v_0/H$ with the range [0.1, 0.4] at intervals of 0.02, resulting in 16 cases of sinusoidal oscillatory shear under different loading amplitudes (${\dot{\gamma}}_0=[0.10, 0.12,..., 0.38, 0.40]$). The 16 datasets are used to train the data-driven framework. We construct a library of candidate functions $\left[\tau,\ \gamma,\dot{\gamma},\ldots,\ddot{\gamma},\tau\gamma,\tau\dot{\gamma},\tau^2\dot{\gamma},\ldots\right]$, encompassing 23 terms that incorporate shear stress $\tau$, shear strain $\gamma$, and their polynomial variations to reconstruct the ODE representing the temporal derivatives of shear stress $\dot{\tau}$. The training process is completed through 10 cycles of the ADO algorithm, where each loop consisted of 1,000 iterations of the Adam optimizer and 100 iterations of the STRidge optimizer. We simultaneously feed data representing various loading conditions characterized by different ${\dot{\gamma}}_0$ amplitudes (at an angular frequency of 0.2$\pi$ rad/s) to recover the model. The dataset is split into training and validation sets in a ratio of 0.8:0.2, which helps to monitor whether the neural network is at risk of overfitting.

Figure~\ref{Fig_PINN_SR_results} illustrates the training process, in which Figs.~\ref{Fig_PINN_SR_results}a and \ref{Fig_PINN_SR_results}b depict the evolution of the coefficients for the 23 candidate functions used to construct the governing equation, with colors representing coefficient values. During the pre-training phase, we employ the L-BFGS algorithm for global optimization without thresholding the coefficients, resulting in all 23 candidate terms being active, leading to a dense and non-parsimonious equation. In contrast, during each ADO iteration, the STRidge method is used to select significant candidate functions, deactivating those with coefficients below a certain threshold by setting them to zero. After three iterations of ADO, the identified equation terms and their coefficients stabilized, with $\frac{d\gamma}{dt}, \mu\left|\frac{d\gamma}{dt}\right|$ and ${\mu}^2\frac{d\gamma}{dt}$ emerging as the most significant terms in the final governing equation. The total loss reduction processes are briefly illustrated in Figs.~\ref{Fig_PINN_SR_results}c and \ref{Fig_PINN_SR_results}d, where it is observed that the total error consistently decreases and eventually stabilizes. More details can be found in \textcolor{red}{Section 2.1.1 of the SI}: during the training process of PINNs, both the DNN loss and the physical loss gradually decrease. Similarly, in sparse regression, both losses reach a stable state, indicating that a unique form of the equation has been identified. Ultimately, a highly concise equation is obtained, as shown in Eq. (\ref{eq7_discovered_equation}):  
\begin{equation}
\frac{d\mu}{dt}=C_1\cdot\frac{d\gamma}{dt}+C_2\cdot\mu\left|\frac{d\gamma}{dt}\right|+C_3\cdot\mu^2\frac{d\gamma}{dt}.\label{eq7_discovered_equation}
\end{equation} with coefficients $C_1=-5.98$, $C_2=-28.05$, and $C_3=-32.41$.  Specifically, the rate of change of friction ($\frac{d\mu}{dt}$) is balanced by three contributions: a linear response term ($C_1\cdot\frac{d\gamma}{dt}$) driven by the shear rate, a dissipative term ($C_2\cdot\mu\left|\frac{d\gamma}{dt}\right|$) coupling the current friction to the shear rate intensity, and a nonlinear response regulatory term ($C_3\cdot\mu^2\frac{d\gamma}{dt}$) that captures the self-driven evolution of friction. The physical implications of each term are described in detail in \textcolor{red}{Section 2.2.3 of SI}. Moreover, these three terms in the equation can be interpreted in terms of micro-mechanisms through the fabric tensor, with details provided in \emph{Discussion}.

We rigorously validate the governing equation with untrained loading scenarios. Figures~\ref{Fig_PINN_SR_results}e and \ref{Fig_PINN_SR_results}f illustrate validations under previously unseen sinusoidal loading conditions, with dashed lines representing DEM results and solid lines depicting the results from the Runge-Kutta integration of the identified equation. Remarkably, the data-driven discovered Eq. (\ref{eq7_discovered_equation}) successfully predicts both the temporal evolution of $\mu$ and the stress-strain response of the material. An additional compelling validation is depicted in Figs.~\ref{Fig_PINN_SR_results}h and \ref{Fig_PINN_SR_results}i: although our governing equation is derived from sinusoidal loading scenarios, we discover its applicability seamlessly extends to Heaviside loading conditions. This represents a distinctly different loading mode, characterized by sudden directional changes, yet Eq. (\ref{eq7_discovered_equation}) adeptly captures the complex dynamic response of the granular system under such discontinuous loadings. This underscores the robust generalizability of the equation.

\subsection{Learning equation coefficients}\label{subsec2-4}

\begin{figure}[htbp]
    \centering
    \includegraphics[width=1\linewidth]{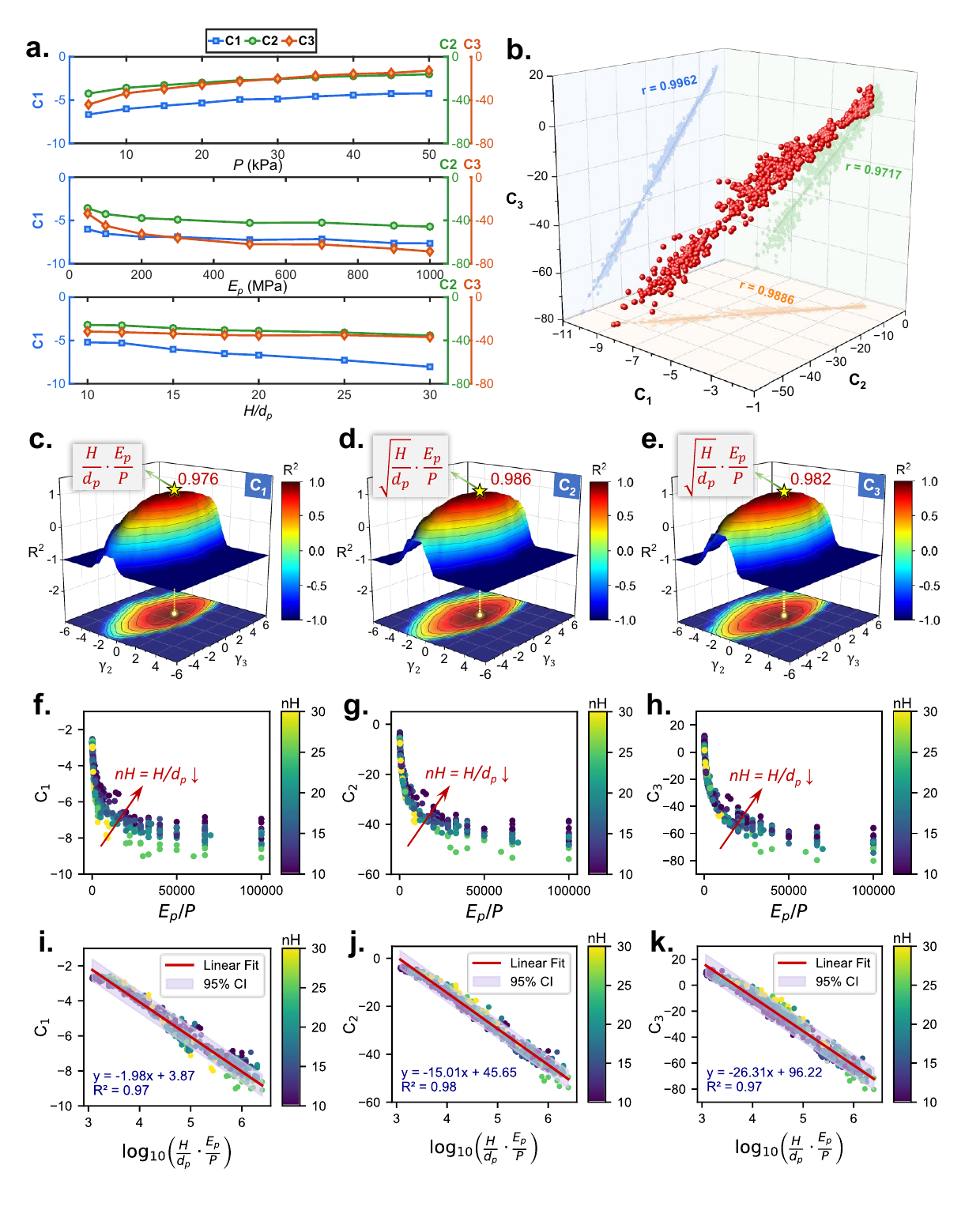}
    \caption{\textbf{Discovery of dominant dimensionless coefficients for governing equation.} \textbf{a} Equation's coefficients vary with the basic quantities of the granular system under oscillatory shearing. \textbf{b} Linear relationship among the coefficients. \textbf{c-e} Dimensionless space constructed with $\gamma_2$ and $\gamma_3$ as coordinates for $C_1$, $C_2$, and $C_3$. The $R^2$ scores denote the best learning performance of the corresponding dimensionless numbers within the space. $R^2$ values less than -1 are displayed as -1. \textbf{f-h} Coefficients' relationship with the primary variable $E_p/{P}$, where color denotes the value of $H/d_p$. \textbf{i-k} Remarkable linear relationship between each coefficient and its dominant dimensionless number on a logarithmic scale, with most falling within the 95\% confidence interval.}
    \label{Fig_Dimensionless_results}
\end{figure}

\begin{sidewaystable}[htbp]
    \centering
    \caption{System variables of the oscillatory simple shear}
    \label{table_systemVariables}
    \begin{tabular}{lll}
        \toprule
        System Variable & Value & Unit \\
        \midrule
        $P$ & 5, 6, 8, \textcolor{red}{\textbf{10}}, 15, 20, 25, 30, 35, 40, 45, 50, 80, 100, 200, 300, 400, 500 & \si{\kilo\pascal} \\
        $H$ & $(10, 12, \textcolor{red}{\textbf{15}}, 18, 20, 25, 30) \cdot d_p$ & \si{\milli\metre} \\
        $T$ & 6, 8, \textcolor{red}{\textbf{10}}, 12, 14, 16, 18, 20 & \si{\second} \\
        $d_p$ & 3, 4, \textcolor{red}{\textbf{5}}, 6, 7, 8, 9, 10 & \si{\milli\metre} \\
        $\rho$ & 2350, 2400, 2450, \textcolor{red}{\textbf{2500}}, 2550 & $\si{\kilogram}\hspace{0.1em}\cdot\hspace{0.1em}\si{\per\cubic\metre}$ \\
        $E_p$ & \textcolor{red}{\boldmath{$50$}}, $100$, $200$, $300$, $500$, $700$, $1000$ & \si{\mega\pascal} \\
        \bottomrule
    \end{tabular}
    \footnotetext{Note: values in red and bold are the representative configurations for the case-specific governing equation discovery.}
\end{sidewaystable}

In this section, we aim to develop a general governing equation capable of describing scenarios across various system variables, rather than being confined to case-specific instances. Intriguingly, the PINN-SR framework consistently yields terms $\dot{\gamma}$, $\mu\left|\dot{\gamma}\right|$, and ${\mu}^2\dot{\gamma}$ , albeilt with varying coefficients, when altering system parameters (e.g. loading conditions, setup height, and particle properties). Figure~\ref{Fig_Dimensionless_results}a demonstrates the influence of variations in key factors on these coefficients. Here, we conduct an in-depth examination of system parameters including normal stress ($P$), height of setup ($H$), sinusoidal shear loading period ($T$), particle density ($\rho$), particle diameter ($d_p$) and Young's modulus of particles ($E_p$), with the aim of establishing a relationship from system parameters $\left\{P,\ H,T,\rho,d_p,E_p\right\}$ to $\left\{C_1,C_2,C_3\right\}$. Detailed results of the search for each individual parameter are provided in \textcolor{red}{Section 3.3 of the SI}. To expedite the investigation of the interplay and expression forms of these parameters, we randomly select variables from a wide range (as indicated in Table~\ref{table_systemVariables}) and conduct extensive DEM simulations. A post-training process is implemented to fine-tune the coefficients of the identified ODE structure. Furthermore, we observe that incorporating more than four distinct data sets of loading amplitude datasets for equation discovery yields diminishing returns in predictive precision, as indicated by the plateau in mean squared error (MSE) (see \textcolor{red}{Section 3.2 of the SI}). Thus, four representative ${\dot{\gamma}}_0$ groups corresponding to the value of the inertial number $I_0=[2.5, 5.0, 7.5, 10]\times10^{-4}$ are employed in all cases to optimize both computational efficiency and model performance for the identification of the governing equations. The results presented here are based on 625 groups of $\left\{C_1,C_2,C_3\right\}$, totaling 2,500 DEM simulations.

Figure~\ref{Fig_Dimensionless_results}b presents an overview of the relationships among $C_1$, $C_2$ and $C_3$ in a three-dimensional space, revealing a remarkable linear correlation, with pairwise correlation coefficients exceeding 0.97. Consequently, when reverting to a steady state, the solutions of Eq. (\ref{steady_dequation}) remain remarkably consistent in varying conditions, indicating that the static friction coefficient $\mu_s$ is minimally affected by these system parameters.

We observe that $C_1$, $C_2$, and $C_3$ are all non-dimensional variables. Thus, we employ the dimensional learning method to discern a low-dimensional and scale-invariant scaling law that optimally encapsulates the $C_i\ (i=1,2,3)$. We assume that each coefficient is governed by a single dominant dimensionless number, as shown in Eqs.~(\ref{eq9_dimensionless_number}) and (\ref{eq10_scaling_law}):
\begin{equation}
\Pi_i=P^{w_{i1}}\cdot d_p^{w_{i2}}\cdot\rho^{w_{i3}}\cdot E_p^{w_{i4}}\cdot T^{w_{i5}}\cdot H^{w_{i6}}.\label{eq9_dimensionless_number}
\end{equation}
\begin{equation}
C_i=f_i(\Pi_i).\label{eq10_scaling_law}
\end{equation}
Here, $i$ represents the $i$th index of the equation term ($i=1,2,3$ in our case); $C_i$ indicates the coefficient of each term (as the output); $\Pi_i$, $f_i$ denote the predominant dimensionless number and the scaling law, respectively; $\boldsymbol{w_i}=\left[w_{i1},\ldots,w_{i6}\right]^T$ denotes the undetermined exponents required for calculating the dimensionless numbers, called dimension vector.

To incorporate dimensional invariance, we conduct a dimensionless analysis, which necessitates that the powers $\boldsymbol{w_i}=\left[w_{i1},\ldots,w_{i6}\right]^T$ adhere to a linear equation system (see Eq. (\ref{eq11_Dw_0})).

\begin{equation}
\boldsymbol{Dw_i}=0.\label{eq11_Dw_0}
\end{equation}
Here, $\boldsymbol{D}$ represents the dimension matrix corresponding to the input variables, as shown in Eq. (\ref{eq12_dimensional_matrix}).
\begin{equation}
\boldsymbol{D}=\left[\begin{matrix}-1&1&-3&-1&0&1\\-2&0&0&-2&1&0\\1&0&1&1&0&0\\\end{matrix}\right]\begin{matrix}\left[L\right]\\\left[T\right]\\\left[M\right]\\\end{matrix}.\label{eq12_dimensional_matrix}
\end{equation}

Within our system, only three basic dimensions, denoted as $\left[L\right]$, $\left[T\right]$, and $\left[M\right]$, exist. All other dimensions in consideration emerge as monomial power functions of the basic dimensions. Eq. (\ref{eq12_dimensional_matrix}) succinctly represents the physical dimensions of system variables, with each column from left to right representing the physical quantities $\left\{P,d_p,\rho,E_p,T,H\right\}$. Eq. (\ref{eq11_Dw_0}) rigorously ensures that the power-law monomials of the output variables in Eq. (\ref{eq9_dimensionless_number}) are rendered dimensionless.

Our current objective is to find an optimal vector $\boldsymbol{w}$ for determining predominant dimensionless numbers and establishing appropriate scaling laws. However, the linear system described by Eq. (\ref{eq11_Dw_0}) is underdetermined, possessing more unknowns than equations, resulting in an infinite number of solutions. To further reduce the dimensionality for machine learning algorithm, inspired by Buckingham's $\pi$ theorem, a strategic way involves expressing as a linear combination of basis vectors, as presented in Eq. (\ref{eq13_basisvector_form}):
\begin{equation}
\boldsymbol{w_{i}}=\gamma_{i1}\boldsymbol{w}_{b1}+\gamma_{i2}\boldsymbol{w}_{b2}+\gamma_{i3}\boldsymbol{w}_{b3}.\label{eq13_basisvector_form}
\end{equation}
where $\boldsymbol{\gamma_{i}}=\left[{\gamma_{i1},\ \gamma}_{i2},\gamma_{i3}\right]$ denotes the set of coefficients associated with the basis vectors, which remains unknown to be determined. In our context, the count of basis vectors corresponds to the difference between the number of system variables (six) and the rank of the dimension matrix (three), as determined by Eq. (\ref{eq12_dimensional_matrix}). In this instance, three basis vectors are detailed in Eqs.~(\ref{eq14_bvector1})-(\ref{eq16_bvector3}):

\begin{equation}
w_{b1}=\left[-1,\ 0,\ 0,\ 1,\ 0,\ 0\right]^T,\label{eq14_bvector1}
\end{equation}
\vspace{0pt} 
\begin{equation}
w_{b2}=\left[0.5,\ -1,\ -0.5,\ 0,\ 1,\ 0\right]^T,\label{eq15_bvector2}
\end{equation}
\vspace{0pt}
\begin{equation}
w_{b3}=\left[0,\ -1,\ 0,\ 0,\ 0,\ 1\right]^T.\label{eq16_bvector3}
\end{equation}

Once $\boldsymbol{\gamma_{i}}$ is determined, the dimensionless form can be uniquely established through (\ref{eq13_basisvector_form}) and (\ref{eq10_scaling_law}). To determine the basis coefficients based on the datasets, a representation learning model is required to approximate $f_i\left(\cdot\right)$. Here, to balance computational efficiency, instead of employing more complex machine learning models, we simply utilize twelfth-order polynomial functions, as shown in Eq. (\ref{eq17_12order_polynomial}). Subsequent results also indicate that polynomial functions have demonstrated a high prediction accuracy.
\begin{equation}
C_i=\beta_{i0}+\beta_{i1}\Pi_i+\beta_{i2}\Pi_i^2+\beta_{i3}\Pi_i^3+\cdots+\beta_{i12}\Pi_i^{12}.\label{eq17_12order_polynomial}
\end{equation}
where $\boldsymbol{\beta_{i}}=\left[\beta_{i0},\beta_{i1},\beta_{i2},{\ldots,\beta}_{i12}\right]^T$ encapsulates the scaling relationship between $C_i$ and $\Pi_i$. To ascertain $\boldsymbol{\gamma_{i}}$ and $\boldsymbol{\beta_{i}}$, we employ an iterative two-level optimization approach as described in the literature \cite{xie2022data}; further details are provided in \textcolor{red}{Section 1.2.1 of the SI}. We set $\gamma_{i1}=1$ to effectively avoid the identification of equivalent dimensionless numbers with different powers and reduce the computational cost. On this basis, we conduct a grid search for $\gamma_{i2}$ and $\gamma_{i3}$ within the range of -6 to 6, with 250 grid points set for each basis coefficient. In the dimensionless space shown in Fig.~\ref{Fig_Dimensionless_results}c-e, for each set of dimensionless numbers $\Pi$ represented by $[\gamma_{i1},\ \gamma_{i2},\ \gamma_{i3}]$, based on the dataset divided into an 80\% training set and a 20\% test set, we use the training-set data to fit the polynomial coefficients $\beta$ according to Eq. (\ref{eq17_12order_polynomial}), thereby determining the specific expression of $f_i\left(\Pi_i\right)$. To evaluate the overall performance of $f_i\left(\Pi_i\right)$, we select the coefficient of determination ($R^2$) of the test set as the indicator. For the coefficient $C_1$ (Fig.~\ref{Fig_Dimensionless_results}c), we uniquely determine the position where $R^2$ reaches its maximum value (0.977), marked as a yellow five-pointed star. At this point, $\gamma_2=0$ and $\gamma_3=1$. Similarly, for the coefficients $C_2$ and $C_3$ (Figs.~\ref{Fig_Dimensionless_results}d and \ref{Fig_Dimensionless_results}e), the maximum values of $R^2$ are 0.984 and 0.976 respectively, and the corresponding coordinates are $\gamma_2=0$ and $\gamma_3=0.5$. Using these optimized basis coefficients, the expression of the dominant dimensionless numbers can be finally determined as:
\begin{equation}
\Pi_1=\frac{H}{d_p}\cdot\frac{E_p}{P}.\label{eq18_dl_number1}
\end{equation}
\begin{equation}
\Pi_2=\Pi_3=\ \sqrt{\frac{H}{d_p}}\cdot\frac{E_p}{P}.\label{eq19_dl_number2}
\end{equation}

Our analysis reveals that both dimensionless numbers are strongly related to $E_p/P$ and $H/d_p$ among all the parameters explored. The prominence of these two terms is justified by the need to account for finite-size and stiffness effects in quasi-static shear~\cite{liu2018size,silbert2007rheology}. Although $\sqrt{\frac{H}{d_p}}\frac{E_p}{P}$ is the dominant dimensionless number for $C_2$ and $C_3$, the $R^2$ values for $\frac{H}{d_p}\frac{E_p}{P}$ remain high at 0.983 and 0.973. To streamline the results, we define $\chi=\frac{H}{d_p}\frac{E_p}{P}$, and the resulting governing equation takes the general form: 
\begin{equation}
\frac{d\mu}{dt}=f_1(\chi)\cdot\frac{d\gamma}{dt}+f_2(\chi)\cdot\mu\left|\frac{d\gamma}{dt}\right|+f_3(\chi)\cdot\mu^2\frac{d\gamma}{dt},\label{final_equation}
\end{equation} where $f_i(\chi)=a_i\log_{10}(\chi)+ b_i$ for $i = 1,2,3$, with coefficients given by $a_1=-1.98, b_1=3.87$; $a_2= -15.01, b_2=45.65$; and $a_3=-26.31, b_3=96.22$.

Figures~\ref{Fig_Dimensionless_results}f-\ref{Fig_Dimensionless_results}h offer a comprehensive depiction of the logarithmic relationships between the coefficients $\left\{C_1,C_2,C_3\right\}$ and the ratio $E_p/P$, as well as the parameter $H/d_p$ (indicated by color). As the ratio $E_p/P$ increases, the coefficients approach saturation, with plateau levels governed by $H/d_p$. Further analysis in logarithmic coordinates reveals robust linear correlations between key dimensionless numbers and $C_i$, with almost all data points falling within the $95\%$ confidence interval of the linear fit (as shown in Figs.~\ref{Fig_Dimensionless_results}i-\ref{Fig_Dimensionless_results}k).

\subsection{Analyzing steady and transient states}\label{subsec2-5}

\textbf{Analyzing steady state.}
The discovered governing equation of the effective friction coefficient ($\mu$) allows us to analyze both the steady and time-dependent states of the granular system. Upon reducing our equation to the steady-state condition, defined by $\frac{d\mu}{dt}=0$, results in Eq. (\ref{eq7_discovered_equation}) degenerates in Eq. (\ref{steady_dequation}) as follows:

\begin{equation}
C_1+C_2\cdot\mu\cdot\mathrm{sgn}\left(\frac{d\gamma}{dt}\right)+C_3\cdot\mu^2=0.\label{steady_dequation}
\end{equation}
In this state, stress is invariant to the strain rate and depends solely on its direction. This observation is consistent with the established theoretical understanding of transient flow within the quasi-static regime, where stress is shear-rate-independent. Thus, our model aligns seamlessly with current theoretical knowledge.

\begin{figure}[htbp]
    \centering
    \includegraphics[width=0.95\linewidth]{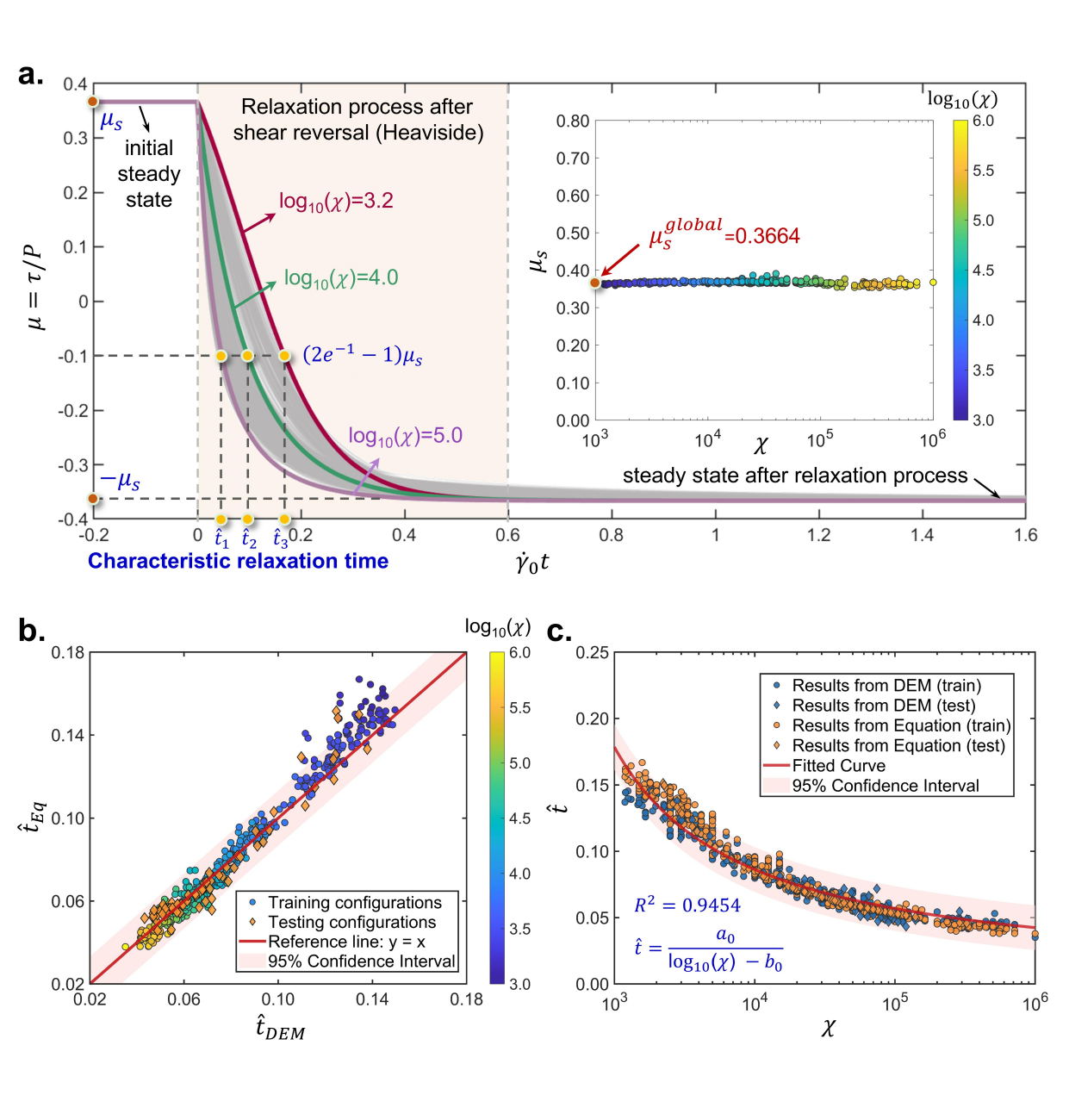}
    \caption{\textbf{One-to-one relationship between the identified dimensionless number and characteristic relaxation time.} \textbf{a} Definition of critical relaxation time. \textbf{b} Relaxation time results based on Eq.~(\ref{relaxation_time}) and DEM simulations. \textbf{c} Fitting results of critical relaxation time in terms of the identified dimensionless number.}
    \label{RelaxationTime}
\end{figure}

Furthermore, solving the quadratic equation Eq.~(\ref{steady_dequation}) provides the steady-state solution ($\mu_s$) for Eq.~(\ref{eq7_discovered_equation}). Our analysis reveals that the $\mu_s$ values form an approximately planar distribution in the three-dimensional $(C_1, C_2, C_3)$ parameter space, as illustrated in \textcolor{red}{Section 2.2.2 of the SI}. Fitting this plane using the form of Eq.~(\ref{steady_dequation}) yields a global $\mu_s$ value, denoted $\mu_s^\text{global} = 0.3664$, which is consistent with the results from existing literature using similar DEM simulation parameters~\cite{jing2022drag}. As a steady-state physical quantity derived from the governing equation, $\mu_s$ is determined by the equation coefficients $(C_1, C_2, C_3)$, which are governed by the identified dimensionless number $\chi$. However, $\mu_s$ exhibits negligible sensitivity to variations in $\chi$ and remains approximately constant (see the inset of Fig~~\ref{RelaxationTime}a). This $\mu_s$ can be related to the well-known $\mu(I)$ rheology, which describes the static friction coefficient as the inertial number approaches zero ($I \to 0$). 

\textbf{Analyzing transient state.}
Under time-dependent loading conditions, the granular system resides in a transient state where the relaxation process is critical. For simplicity, we examine a Heaviside scenario: the relaxation process following instantaneous shear rate reversal, switching from $\dot\gamma_0$ to $-\dot\gamma_0$ (with $\dot\gamma_0>0$). \textcolor{red}{Section 2.3.2 of the SI} displays the DEM simulation results for the same configuration under varying amplitudes $\dot\gamma_0$, revealing that their relaxation processes exhibit consistency in dimensionless time $\dot\gamma_0t$. This is manifested by all curves collapsing on a single curve during the relaxation period after the sudden shear reversal at $t=0$. Herein, we analyze this relaxation process via our proposed equation, which is mathematically a Riccati equation. Incorporating boundary conditions yields an analytical solution in sigmoid form, as shown in Eq.~(\ref{muAnalytical_solution}). Detailed derivations are provided in \textcolor{red}{Section 2.3.1 of the SI}.

\begin{align}
\mu(t)&=\mu_s+\frac{C_2 - 2C_3\mu_s}{C_3-\frac{C_2}{2\mu_s}\exp[-\dot{\gamma}_0(C_2 - 2C_3\mu_s)t]}.\label{muAnalytical_solution}
\end{align}

\begin{align}
\mu(t)&=\mu_2-(\mu_2-\mu_1)\cdot e^{-t_0/t}.\label{KWW_law}
\end{align}

From Eq.~(\ref{muAnalytical_solution}), we plot configuration-dependent curves (see Fig.~\ref{RelaxationTime}a), with three representatives with $\log_{10}(\chi)$ = 3.2, 4.0, and 5.0. To characterize the relaxation process, we introduce a dimensionless characteristic relaxation time $\hat{t}$. Following the stretched exponential Kohlrausch-Williams-Watts law (Eq.~(\ref{KWW_law}), consider $t=t_0$, $\mu_1=\mu_s$, and $\mu_2=-\mu_s$), $\hat{t}$ is the dimensionless time at which $\mu=(2e^{-1}-1)\mu_s$. This yields the analytical expression for $\hat{t}$ in Eq.~(\ref{relaxation_time}).

\begin{equation}
\hat{t}=\frac{1+\ln\left(C_{2}-2(1-e^{-1})\frac{C_{3}\mu_{s}}{C_{2}}\right)}{|C_{2}-2C_{3}\mu_{0}|}.\label{relaxation_time}
\end{equation}

Figure~\ref{RelaxationTime}b compares the characteristic relaxation times from DEM simulations ($\hat{t}_{DEM}$) and from Eq.~(\ref{relaxation_time}) ($\hat{t}_{Eq}$). All data points are close to the line $y=x$, confirming the validity of the equation identified from sinusoidal cases to capture the relaxation characteristics of the Heaviside case.We further examine a test configuration with $d_p=15~\text{mm}$, which is completely outside of the system parameters listed in Table~\ref{table_systemVariables}. Using Eq.~(\ref{final_equation}) to derive the governing equation coefficients and subsequently Eq.~(\ref{relaxation_time}) to compute $\hat{t}$, these testing points also fall close to $y=x$, demonstrating that Eq.~(\ref{final_equation}) possesses robust generalization and extrapolation capabilities in different configurations. Figure~\ref{RelaxationTime}c has shown the one-to-one relationship between the identified dimensionless number ($\chi$) and characteristic relaxation time ($\hat{t}$). At times, in the absence of a governing equation where only system parameters are available, the findings of this study can still be applied to predict the relaxation characteristics of the system. Herein, we recommend employing an empirical hyperbolic formula $\hat{t}=a_0/(\log_{10}(\chi)-b_0)$, where $a_0=0.167$ and $b_0=2.065$. From this, we conclude that the characteristic relaxation time increases with smaller $H/d_p$, softer particles (lower $E_p$) and higher pressure $P$. Notably, this denotes the dimensionless relaxation time ($\dot{\gamma}_0 t$), distinct from physical time ($t$).

We first elaborate on the role of $H/d_p$, which is the number of layers of granular flows. For smaller $H/d_p$, during transient evolution, the force chain scale ($L_{fc}$) is comparable to $H$ or even spans the entire system, yielding greater strength (manifested as higher viscosity). Conversely, for larger $H/d_p$, $L_{fc}$ is smaller than $H$, inter-particle force chains break and rearrange more easily to facilitate the system's quickly reaching a new equilibrium. \textcolor{red}{Section 2.4.1 of the SI} provides a detailed comparison of force chain evolution in systems with $H/d_p=10$ and $50$. These findings are consistent with previous reports that reduced system size leads to enhanced material strengthening~\cite{liu2018size}. 

Furthermore, the second ratio $E_p/P$ relates to the contact stiffness number ($\kappa$). For Hertzian contacts, $\kappa=(E/P)^{2/3}$ as defined in the reference~\cite{da2005rheophysics}. Within $\mu(I)$ rheology, $\kappa$ is typically assumed large ($\kappa>10^{4}$), making the inertial number the only parameter that dictates the rheology. Our observations corroborate that the effect of $E_p/P$ saturates at large values, as shown in Figs.~\ref{Fig_Dimensionless_results}f–h. However, $\kappa$ becomes highly significant in low-inertial number regimes, particularly when $\kappa>10^4$ is not met~\cite{duanyifei2019}. In the quasi-static regime, $\kappa$ modulates contact lifetimes and thereby governs rheological behaviors~\cite{silbert2007rheology}. Studies have shown that $\kappa$ is proportional to the square of the time scale ratio $(\tau_i/\tau_c)^2$, where $\tau_i=d/\sqrt{P/\rho}$ denotes the inertial time and $\tau_c$ is the collision time~\cite{da2005rheophysics}. Our parameter range, $E_p/P\leq10^5$, corresponds to $\kappa\leq10^{10/3}$. For high $\kappa$ (high $E/P$), $\tau_c$ is much smaller than $\tau_i$, resulting in rapid structural reorganization dominated by inertia. In contrast, for low $\kappa$ (low $E/P$), $\tau_c\sim \tau_i$. This reduced contact stiffness prolongs the collision time, slows the transmission of interparticle force, delays the reorganization of the force chain network, and ultimately increases the system relaxation time.

In summary, our identified dimensionless number $\chi$ encapsulates both the size dependence and the influence of microscale competition between the two time scales ($\tau_c$ and $\tau_i$) in the transient state. The change of $\mu$ will be faster under time-dependent shearing when $\chi$ is larger.

\section{Discussion}\label{sec3}
\subsection{Microscopic interpretation via fabric tensor}\label{sec:Microscopic_interpretation}

Previous rheological laws for granular matter are typically formulated solely as constitutive relations rather than as evolution equations. In this study, we derive a governing equation that captures the evolution of the effective friction coefficient ($\mu$), allowing us to gain a deeper understanding of transient granular physics. This section discusses the physical interpretations of the equation terms and clarifies the associated microscopic mechanisms.

Intriguingly, we find that the overall structure of our identified equation is reminiscent of an earlier microstructural equation that describes the evolution of the fabric tensor in the literature~\cite{sun2011constitutive}. Separately, a related study~\cite{azema2014internal} demonstrates that $\mu$ originates from three distinct components associated with contact network and force transmission: (i) contact anisotropy, (ii) force chain anisotropy, and (iii) friction mobilization. This investigation also indicates that contact anisotropy predominates in the quasi-static regime, which prompts us to investigate the physical interpretation of each term in the equation from the perspective of fabric tensors.

\begin{figure}[htbp]
    \centering
    \includegraphics[width=0.99\linewidth]{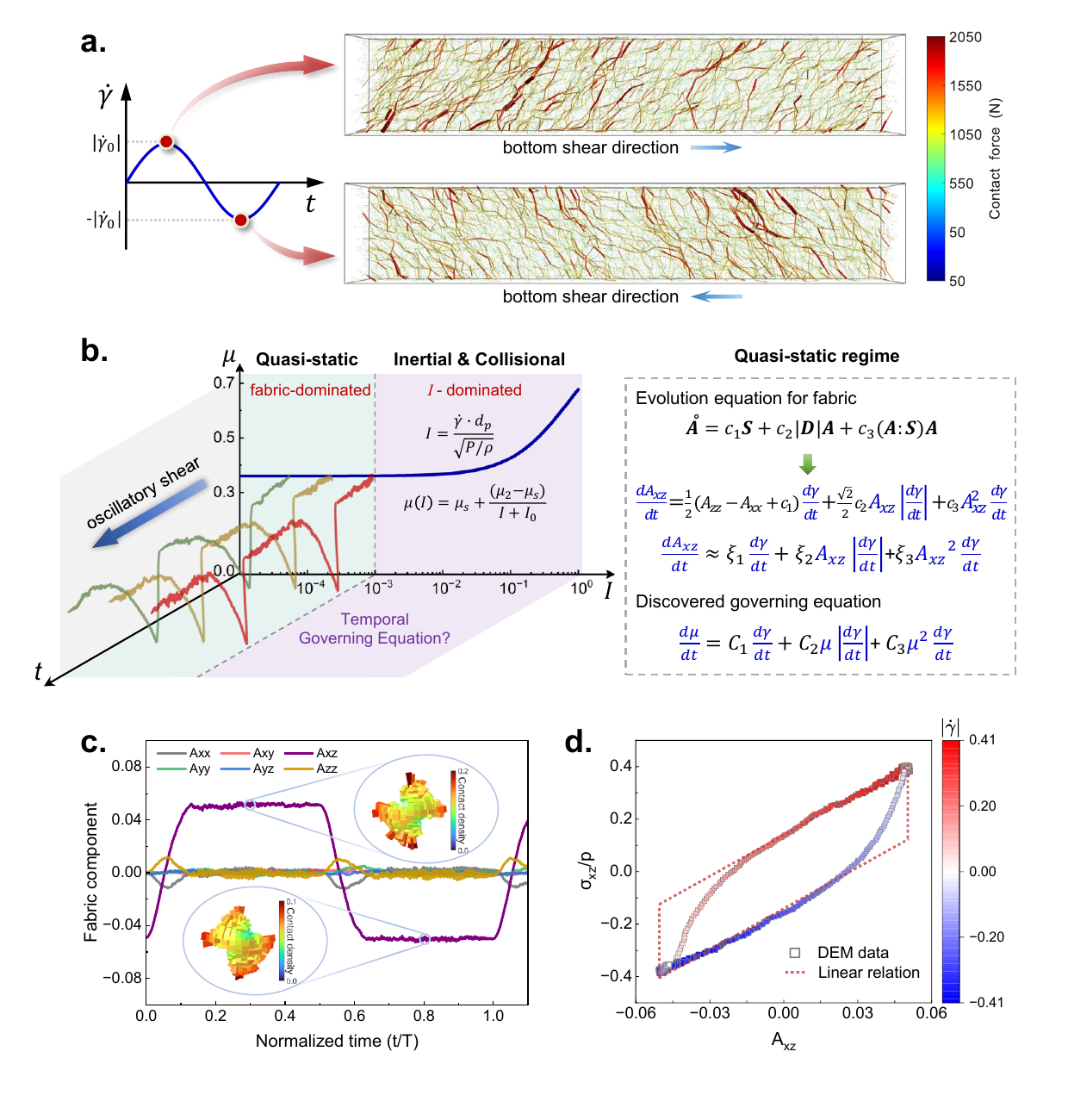}
    \caption{\textbf{Microscopic interpretation of the discovered governing equations.} \textbf{a} Evolution of microstructure during oscillatory shear. \textbf{b} Contribution of this work in granular rheology. \textbf{c} Fabric component. \textbf{d} Linear relation between frictional coefficient ($\mu=\sigma_{xz}/{P}$) and diagonal component of fabric tensor ($A_{xz}$).}
    \label{Fig_physical_meaning}
\end{figure}

As depicted in Fig.~\ref{Fig_physical_meaning}a, we observe a notable evolution in the microstructure of granular systems due to shear reversal. The force chain structure exhibits pronounced directionality at peak shear rate ($\left|{\dot{\gamma}}_0\right|$ and $-\left|{\dot{\gamma}}_0\right|$). To characterize microstructure anisotropy, we define the fabric tensor $\mathbf{A}$ as a symmetric traceless tensor of second order (Eq. (\ref{eq20_fabric})). The eigenvectors of $\mathbf{A}$ give the principal directions of the mean contact orientations, while its eigenvalues quantify the degree to which these contacts align with the principal directions. Hence, anisotropy can be assessed by the disparity between the largest and smallest eigenvalues.
\begin{equation}
\mathbf{A}=\frac{1}{N_C}\sum_{i=1}^{N_C}{\mathbf{n}^i\mathbf{n}^i-\frac{1}{3}\mathbf{I}}.\label{eq20_fabric}
\end{equation}
Here, $\boldsymbol{n^i}$ denotes the $i$-th unit contact vector among the $N_C$ contacts in the granular assembly, while $\boldsymbol{I}$ represents the unit tensor. An equation governing the evolution of the fabric tensor has been postulated \cite{sun2011constitutive}, as illustrated:
\begin{equation}
\overset{\circ}{\operatorname*{\operatorname*{\mathbf{A}}}}=c_1\mathbf{S}+c_2|\mathbf{D}|\mathbf{A}+c_3(\mathbf{A}{:}\mathbf{S})\mathbf{A}.\label{eq21_evolution_equation_fabric}
\end{equation}
Here, $\overset{\circ}{\operatorname*{\operatorname*{\mathbf{A}}}}$ is the Jaumann Rate of $\mathbf{A}$, defined as $\overset{\circ}{\operatorname*{\operatorname*{\mathbf{A}}}}=\dot{\mathbf{A}}+\mathbf{A}\cdot\mathbf{W}-\mathbf{W}\cdot\mathbf{A}$, where $\mathbf{W}$ is the spin tensor given by $\mathbf{W}=\frac{1}{2}\left(\mathbf{\nabla}\boldsymbol{v}-\left(\mathbf{\nabla}\boldsymbol{v}\right)^T\right)$ with $\boldsymbol{v}$ denoting the velocity field, and $\dot{\mathbf{A}}$ signifies its material time derivative. The strain-rate tensor and deviatoric strain-rate tensor are defined as $\mathbf{D}=\frac{1}{2}\left(\mathbf{\nabla}\boldsymbol{v}+\left(\mathbf{\nabla}\boldsymbol{v}\right)^T\right)$ and $\mathbf{S}=\mathbf{D}-\frac{1}{3}tr\left(\mathbf{D}\right)\mathbf{I}$, respectively. And $\left|\cdot\right|$ represents the modulus of tensor, thus, $\left|\mathbf{D}\right|=\sqrt{\mathbf{D}:\mathbf{D}}$. $c_1\sim c_3$ (distinct from the capital $C_1\sim C_3$) in Eq. (\ref{eq7_discovered_equation}) are simply defined as constants in the theoretical paper \cite{sun2011constitutive}.

Notablely, in our DEM simulations, the particle velocities in the $y$ and $z$ directions are essentially noise around zero, lead to $\mathbf{D}\approx\mathbf{S}\approx\frac{1}{2}\dot{\gamma}\left(\hat{\mathbf{i}}\otimes\hat{\mathbf{j}}+\hat{\mathbf{j}}\otimes\hat{\mathbf{i}}\right)$. This implies that, despite Eq. (\ref{eq21_evolution_equation_fabric}) being a tensor-based governing equation, its structure closely resembles that of Eq. (\ref{eq7_discovered_equation}). Moreover, both our DEM results and data from reference \cite{sun2011constitutive} illustrate that the values of $A_{xx}$ and $A_{zz}$ are significantly smaller than $A_{xz}$, while $A_{yy}$, $A_{xy}$, and $A_{yz}$ are nearly zero, as depicted in a representative DEM simulation case in Fig.~\ref{Fig_physical_meaning}c. Rose diagrams of the 3D contact information at representative moments are also presented in Fig.~\ref{Fig_physical_meaning}c, highlighting pronounced spatial directional evolution. Additionally, we observe that the difference between the largest and smallest eigenvalues closely approximates $2A_{xz}$, serving as an indicator of anisotropy. Hence, our attention is directed towards the degenerated evolution equation for the $A_{xz}$ component, as delineated in Eq. (\ref{eq22_fabric_Axz}), with comprehensive mathematical derivations provided in \textcolor{red}{Section 2.4.3 of the SI}.
\begin{equation}
\frac{dA_{xz}}{dt}=\frac{1}{2}\left(A_{zz}-A_{xx}+c_1\right)\frac{d\gamma}{dt}+{\frac{\sqrt2}{2}c_2}A_{xz}\left|\frac{d\gamma}{dt}\right|+c_3A_{xz}^2\frac{d\gamma}{dt}.\label{eq22_fabric_Axz}
\end{equation}

Our study demonstrates that the contribution of $A_{zz}-A_{xx}$ to the coefficient of $\frac{d\gamma}{dt}$ is consistently below 1\% and thus negligible. When this term is disregarded, Eq. (\ref{eq7_discovered_equation}) and (\ref{eq22_fabric_Axz}) become structurally identical, with the former describing the evolution of $\mu$ and the latter that of $A_{xz}$ (as shown in Fig.~\ref{Fig_physical_meaning}b). Fig.~\ref{Fig_physical_meaning}d quantitatively illustrates the relationship between $\mu$ and $A_{xz}$, which is predominantly linear, with minor deviations occurring within narrow intervals during reversal loading. In particular, a higher absolute value of fabric anisotropy ($|A_{xz}|$) gives rise to a faster evolution of $\mu$ under time-varying shear conditions. By assuming $\mu\approx kA_{xz}$ ($k$ is a constant coefficient characterizing the linear relationship), the formal consistency between Eq. (\ref{eq7_discovered_equation}) and (\ref{eq22_fabric_Axz}) can be elucidated. Consequently, we can interpret the microscopic physical meaning of each term on the right-hand side in Eq. (\ref{eq7_discovered_equation}) from the perspective of the fabric tensor's evolution, as analyzed in reference \cite{sun2011constitutive}: (i) The first term ($\frac{d\gamma}{dt}$) highlights the role of shear in inducing anisotropy; (ii) the second term ($\mu\left|\frac{d\gamma}{dt}\right|$) provides stability to the system, enabling it to reach a steady state even at high strain levels; (iii) the third term (${\mu}^2\frac{d\gamma}{dt}$) exhibits dual characteristics: it initially destabilizes the system following shear reversal, but subsequently contributes to its stability. This dual effect allows for the modulation of the system's behavior.

\subsection{Summary and outlook}

In conclusion, we provide a comprehensive physical interpretation of each term in the data-driven discovered equation, which not only validates the accuracy of our equation extraction using PINNSR-DA from an alternative perspective but also expands the relevant knowledge base. Most importantly, this work represents a pioneering exploration of a novel research paradigm, `Data Mining-Governing Equations-Physical Mechanisms', as opposed to the traditional, slow, and complex derivation process of `First Principles Physics-Governing Equations'.

There are also some potential limitations inherent in the current PINNSR-DA framework for hidden rheology discovery. Firstly, we must acknowledge that PINN-SR incurs significantly higher computational costs in equation extraction compared to the state-of-the-art SINDy method, predominantly attributed to the protracted training process of the deep neural network (DNN). Moreover, distinct from case-specific mining, dimensional learning of the coefficients necessitates multiple datasets, which increases the difficulty and time required for data collection. Nevertheless, the time invested in DNN training and data collection is worthwhile for the development of novel rheological constitutive models, which has already expedited the traditional manual analysis workflow. Secondly, similar to other SINDy-family methods, PINNSR-DA relies on prior knowledge for candidate library construction. An incomplete library makes it difficult to mine suitable equations. Thirdly, while our approach separates the process into two sequential steps (equation-term mining and coefficient-expression determination), the influence of some system parameters may not be fully isolated from equation terms and coefficients. For example, when the loading period is short ($<$15s), although the period information is incorporated into the terms by $\gamma$, it still affects the preceding coefficients. Lastly, whether the current PINNSR-DA can discern a universal governing equation for granular flows spanning quasi-static to inertial regimes remains unconfirmed. Although equation terms identified across these regimes might diverge, a more compelling hypothesis posits an invariant form with mere coefficient variations. This follows from the tripartite formulation (linear response, nonlinear response and energy dissipation), seemingly capable of capturing the balance between all external drives and internal adjustments. However, coefficient expressions in inertial regimes beyond the scope of this study await further exploration.

Despite the aforementioned limitations of the method, we have identified a new dimensionally homogeneous differential equation for steady-state and transient granular flows. To the best of our knowledge, it marks the first successful application of data-driven discovery techniques to the field of unknown governing equations for granular flows, which extends beyond the scope of the $\mu(I)$ rheology and quasi-static soil mechanics model. Fig.~\ref{Fig_physical_meaning}b highlights the important contributions of the equation presented in this paper to the field of granular mechanics. Nonstationary-state issues remain at the forefront of research because of the absence of unified constitutive laws for transient processes. In this work, we extended the $\mu(I)$ rheology model in the quasi-static regime, incorporating the time dimension. The governing equation we derived can degenerate into steady-state form, which aligns with established principles in soil mechanics and $\mu(I)$ rheology for small inertial numbers, and is also applicable to unsteady processes driven by external load variations. Future work will aim to broaden this framework to encompass rate-dependent inertial states and derive the pertinent temporal governing equations.

Moving forward, PINNSR-DA can be utilized to enhance our comprehension of more challenging issues in granular physics, such as jamming, segregation, flow-particle interactions, and other multiphysics phenomena. The governing equations of these issues can be systematically integrated into continuum models for large-scale simulations.

\section{Methods}\label{sec4}
\noindent\textbf{Discrete element method}

For the simulations, we used the discrete element method (DEM) to perform three-dimensional (3D) oscillatory shear test simulations, resolving granular dynamics at the particle scale. DEM explicitly computes the trajectories of individual particles by solving Newton's equations of motion (see Eq.~(\ref{NewtonSecond})). 
\begin{align}
m_i \frac{d^2 \mathbf{r}_i}{dt^2} &= \sum_{j} \mathbf{F}_{ij}^{c} + \mathbf{F}_i^{\text{ext}},~~~I_i \frac{d\boldsymbol{\omega}_i}{dt}= \sum_{j} \mathbf{M}_{ij}.\label{NewtonSecond}
\end{align}
Here, $m_i$, $I_i$, $\mathbf{r}_i$, and $\boldsymbol{\omega}_i$ denote the mass, moment of inertia, position, and angular velocity of the $i\text{th}$ particle, respectively;  $\mathbf{F}_{ij}^{c}$ and $\mathbf{M}_{ij}$ are contact forces and torques; $\mathbf{F}_i^{\text{ext}}$ denotes external forces such as gravity. In this study, since gravity is not considered, we have \(\mathbf{F}_i^{\text{ext}}=0\).

Interparticle and particle-boundary interactions were modeled using the Hertzian contact law~\cite{adams2000contact} for normal forces, paired with Mindlin-Deresiewicz theory~\cite{mindlin1949compliance,mindlin1953elastic} for tangential components, which accounts for frictional effects and contact history.

\begin{itemize}
    \item \textbf{Normal contact force}: The Hertz model describes two spheres of radii $R_i$ and $R_j$, Young's moduli $E_i$ and $E_j$, and Poisson's ratios $v_i$ and $v_j$, as a function of the normal overlap $\delta_n$:
    \begin{align}
    {F}_n=\frac{4}{3}E^*\sqrt{R^*}\delta_n^{3/2},\label{DEM_normalforce}
    \end{align}
    \begin{align}
    E^*=\left(\frac{1-\nu_i^2}{E_i}+\frac{1-\nu_j^2}{E_j}\right)^{-1},~~~R^*=\left(\frac{1}{R_i}+\frac{1}{R_j}\right)^{-1},\label{DEM_effValue}
    \end{align}
    where $E^*$ and $R^*$ denote effective modulus and effective radius, respectively.
    
    \item \textbf{Tangential contact force}: Following Mindlin-Deresiewicz theory, tangential force ${F}_t$ arise from relative tangential displacement $\delta_t$:
    \begin{align}
    {F}_t &= -k_t {\delta}_t,\label{DEM_tangentialforce2}
    \end{align}
    \begin{align}
    k_t &= 8G^* \sqrt{R^* \delta_n},\label{DEM_tangentialforce3}
    \end{align}
    \begin{align}
    \frac{1}{G^*} &= \frac{2-\nu_i}{G_i} + \frac{2-\nu_j}{G_j}.\label{DEM_tangentialforce4}
    \end{align}
    
   Here, $G^*$ is the effective shear modulus. Coulomb friction constrained the magnitude: $\left| {F}_t \right| \leq \hat{\mu}_s F_n$, where $\hat{\mu}_s$ is the interparticle friction coefficient.
\end{itemize}

In addition, normal restitution coefficients $e$ (set to 0.8 for all interactions) are used to model inelastic energy loss during collisions, defined as $e=-v_{\text{post}}^{n}/v_{\text{pre}}^{n}$, where $v_{\text{pre}}^{n}$ and $v_{\text{post}}^{n}$ are the normal components of relative velocity before and after contact.
The equations of motion are integrated using a velocity-Verlet algorithm~\cite{radjai2011discrete} with a time step $\Delta t=1\times10^{-5}$~s, which is less than \text{1\%} of the estimated particle collision time. Additional details regarding the simulation setup can be found in \emph{Data Collection} section.

\vspace{\baselineskip}
\noindent\textbf{Neural network architecture}

The proposed network architectures of PINNSR are depicted in Fig.~\ref{Fig_framework}a. We consider dual inputs: 
the temporal coordinate $\boldsymbol{t} \in \mathbb{R}^{K \times L}$ and the shear rate amplitude $\boldsymbol{\dot{\gamma}_0} \in \mathbb{R}^{K \times L}$. 
Here, $K$ denotes the number of sinusoidal DEM simulations (with $\dot{\gamma}(t)=\dot{\gamma}_0 \sin\left( \frac{2\pi}{T}\cdot t \right)$) featuring distinct $\dot{\gamma}_0$ amplitudes, used for the discovery of case-specific equations. Each simulation yields a data vector of length $L = t_m / \delta_t$, where $t_m$ represents the total measurement duration ($2T$ or $1.2T$ in our study), and $\delta_t = 0.001~\mathrm{s}$ is the sampling interval in DEM simulations. Note that a preprocessing step is applied here: $\boldsymbol{\dot{\gamma}_0} \in \mathbb{R}^{K \times 1}$ is expanded to $\boldsymbol{\dot{\gamma}_0} \in \mathbb{R}^{K \times L}$ to align with $\boldsymbol{t}$ via replication across temporal dimensions. The outputs, $\boldsymbol{\tau}$ and $\boldsymbol{\gamma}$, are latent solutions approximated by fully connected deep neural networks. Note that while the strain response $\boldsymbol{\gamma}$ could be derieved from predefined functions of $\boldsymbol{\dot{\gamma}_0}$, discontinuous or kinked functional forms (e.g., the Heaviside function) hinder the evaluation of high-order derivatives ($\boldsymbol{\ddot{\gamma}_0},\boldsymbol{\dddot{\gamma}_0}\ldots$) at singular points. Incorporating $\boldsymbol{\gamma}$ into the outputs preserves the consistency of the framework and facilitates automatic differentiation for its high-order derivatives. The dataset is split 8:2 into training and testing subsets, yielding $\boldsymbol{t}^{\text{train}}, \boldsymbol{\dot{\gamma}_0}^{\text{train}}, \boldsymbol{\tau}^{\text{train}}, \boldsymbol{\gamma}^{\text{train}}\in\mathbb{R}^{K \times \lceil 0.8L \rceil}$ and $\boldsymbol{t}^{\text{test}}, \boldsymbol{\dot{\gamma}_0}^{\text{test}}, \boldsymbol{\tau}^{\text{test}}, \boldsymbol{\gamma}^{\text{test}}\in\mathbb{R}^{K \times \lfloor 0.2L \rfloor}$, where $\lceil \cdot \rceil$ and $\lfloor \cdot \rfloor$ denotes the ceiling and floor function for integer dimensions.

The architecture of neural networks, particularly the number of hidden layers and neurons per layer, plays a pivotal role in predictive accuracy and algorithmic performance. We find that widths of 15 to 30 neurons and depths of 4 to 8 hidden layers yield optimal stability across datasets from various oscillatory shear configurations. In this study, we adopt 6 hidden layers with 20 neurons each. ReLU and similar locally linear activation functions are excluded as their piecewise linearity nullifies derivatives of order two and higher. Instead, the hyperbolic tangent (tanh) is employed for its retained nonlinearity (supporting high-order differentiation) and unbiased estimation of both positive and negative values. 

The total loss of PINNSR is a sparsity-regularized and physics-constrained function, formulated as Eq.~(\ref{eq4_PINNSR_loss}). The neural network architecture focuses on optimizing the first two terms: data loss $\mathcal{L}_D$ and residual physical loss $\mathcal{L}_P$. The data loss function is defined as:
\begin{equation}
\mathcal{L}_D(\boldsymbol{\theta};\mathcal{D}_m)=\frac{1}{N_m}\left\|\mathbf{Y}^\theta-\mathbf{Y}^m\right\|_2^2.\label{data_loss}
\end{equation}
Here, $\mathbf{Y}^\theta=\{\boldsymbol{\tau}^\theta, \boldsymbol{\gamma}^\theta\}$ denotes the corresponding solution approximated by the neural network, while $\mathbf{Y}^m=\{\boldsymbol{\tau}^m, \boldsymbol{\gamma}^m\}$ represents the measured output data. $N_m$ indicates the total count of measured data points, and $\left\|\cdot\right\|_2$ refers to the Frobenius norm. The residual physical loss is computed at the sampled collocation points $\mathcal{D}_c$ as:

\begin{equation}
\mathcal{L}_P(\boldsymbol{\theta},\boldsymbol{\Lambda};\mathcal{D}_c)=\frac{1}{N_c}\left\|\boldsymbol{\dot{\tau}}(\boldsymbol{\theta})-\boldsymbol{\Phi}(\boldsymbol{\theta})\boldsymbol{\Lambda}\right\|_2^2,\label{physics_loss}
\end{equation}
where $\boldsymbol{\dot{\tau}}$ and $\boldsymbol{\Phi}$ represent the discretized temporal derivative of $\boldsymbol{\dot{\tau}}$ and the candidate library at collocation points, respectively; $N_c$ denotes the total number of collocation points over time. The collocation set $\mathcal{D}_c$ is dynamically adjusted at each training step. Gaussian kernel density estimation (KDE) is applied to the temporal domain with weights proportional to residual physical losses at $\mathcal{D}_m$, updating the new collocation points. These adaptive sampling method prioritize high-loss regions, enhancing sampling denity where physical constraints are less satisfied. The total sample count is defined as $N_c=\vartheta N_m$, where $\vartheta$ is a hyperparameter controlling collocation density. We set $\vartheta=50$ in this study. More details can be found in \textcolor{red}{Section 1.1.2 of the SI}.

\vspace{\baselineskip}
\noindent\textbf{Alternating direction optimization}

We employ the alternating direction optimization (AOD) algorithm from prior work~\cite{chen2021physics} to address the optimization problem defined by Eq.~(\ref{eq4_PINNSR_loss}), as $l_0$ regularization renders it an NP-hard problem. This approach decomposes the problem into two more tractable sub-optimization tasks, as detailed in Eqs.~(\ref{eq5_argmin1}) and (\ref{eq6_argmin2}). Sparse coefficients $\boldsymbol{\Lambda}$ are updated using sequential thresholding ridge regression (STRidge)~\cite{rudy2017data}, followed by training neural network parameters $\boldsymbol{\theta}$ via Adam optimizer in the same iteration. Prior to executing AOD cycles, we pretrain the PINNSR problem by relaxing the $\|\boldsymbol{\Lambda}\|_0$ in Eq.~(\ref{eq4_PINNSR_loss}) to $\|\boldsymbol{\Lambda}\|_1$, then applying the L-BFGS algorithm to brute-force minimize the $\mathcal{L}_{pre}$ as defined in Eq. (\ref{PINNSR_loss_pretrain}). The pretraining step is critical for initialization with pretrained parameters$\{\boldsymbol{\theta}^*,\boldsymbol{\Lambda}^*\}$.

\begin{equation}
\mathcal{L}_{pre}(\boldsymbol{\theta},\boldsymbol{\Lambda};\mathcal{D}_m,\mathcal{D}_c)=\mathcal{L}_D(\boldsymbol{\theta};\mathcal{D}_m)+\lambda\cdot\mathcal{L}_P(\boldsymbol{\theta},\boldsymbol{\Lambda};\mathcal{D}_c)+\xi\cdot\|\boldsymbol{\Lambda}\|_1.\label{PINNSR_loss_pretrain}
\end{equation}

After ADO cycles, we fine-tune the coefficients of the identified governing structures using the L-BFGS algorithm to enhance the accuracy of the equation coefficients.The theory and algorithm flowchart of this method are presented in \textcolor{red}{Section 1.1 of the SI}.

\vspace{\baselineskip}
\noindent\textbf{Two-level machine learning based dimensional analysis}

Two-level machine learning based dimensional analysis is inspired by the prior work~\cite{xie2022data}. This approach has been stated in \emph{Learning the equation coefficients} section. More details and the alogrithm are elaborated in \textcolor{red}{Section 1.2 of the SI}.


\section{Data Availability}\label{sec5}
All the used datasets in this study are available on GitHub at \url{https://github.com/hanxuhku/PINNSR-DA}.

\section{Code Availability}\label{sec6}
All the source codes to reproduce the results in this study are available on GitHub at \url{https://github.com/hanxuhku/PINNSR-DA}.

\bibliography{main}

\section{Acknowledgments}\label{Acknowledgments}
This work was supported by Research Grants Council of Hong Kong [GRF No. 17205821, GRF No. 17205222, and GRF No. 17200724] (C.Y.K.) and the Hong Kong PhD Fellowship Scheme (X.H.). The computations were performed using research computing facilities offered by Information Technology Services, the University of Hong Kong.

\section{Author contributions}\label{Author_contributions}
Y.D.S., L.J., C.Y.K., and G.Y. contributed to the ideation and design of the research. L.J. and X.H. conducted the DEM simulations and collected the rheological datasets. X.H., G.Y. and L.J. developed the data-driven framework. X.H. wrote the codes, the manuscript and supplementary information. L.J., C.Y.K., G.Y., and  Y.D.S. revised the manuscript. Y.D.S. and X.H. contributed to the mathematical derivations. All authors contributed to numerous discussions of microscopic interpretations. C.Y.K. supervised the entire project.

\section{Competing interests}\label{Competing interests}
The authors declare no competing interests.

\section{Additional information}\label{Additional information}
A supplementary video demonstrating the oscillatory shear tests for both sinusoidal and Heaviside inputs is available online at \url{https://www.youtube.com/watch?v=Zzx57moZebE}.
\end{document}